\documentclass[aps,prd,fleqn,superscriptaddress,nofootinbib]{revtex4}
\usepackage{graphicx,color,natbib}
\usepackage{amsmath,amssymb,amsfonts}

\usepackage{epsfig,epsf}
\usepackage{dcolumn}
\usepackage{float}
\usepackage{bm}

\newcommand{\bse}{\begin{subequations}}
\newcommand{\ese}{\end{subequations}}
\newcommand{\be}{\begin{equation}}
\newcommand{\ee}{\end{equation}}
\newcommand{\bea}{\begin{eqnarray}}
\newcommand{\eea}{\end{eqnarray}}
\newcommand{\ba}{\begin{array}}
\newcommand{\ea}{\end{array}}

\input amssym.def
\input amssym.tex

\usepackage[colorlinks=true, linkcolor=blue, bookmarks=true]
{hyperref}

\begin{document}

\title{Holographic subregion complexity in a moving strongly coupled plasma }
\author{Mohammad Mahdi Daryaei Goki\footnote{m\_daryaeigoki@sbu.ac.ir}}
\affiliation{Department of Physics, Shahid Beheshti University, Tehran, Iran}
\author{Mohammad Ali-Akbari\footnote{m\_aliakbari@sbu.ac.ir}}
\affiliation{Department of Physics, Shahid Beheshti University, Tehran, Iran}
\author{Mahsa Lezgi\footnote{mahsalezgee@yahoo.com}}
\affiliation{School of Physics, Institute for Research in Fundamental Sciences (IPM), 19538-33511, Tehran, Iran}
\author{Vahid Esrafilian\footnote{v\_esrafilian@sbu.ac.ir}}
\affiliation{Department of Physics, Shahid Beheshti University, Tehran, Iran}

\begin{abstract}
We study holographic subregion complexity in a moving strongly coupled plasma in dimensions $d=2,3,4$, which is holographically dual to a boosted black brane metric in a higher dimensional geometry. The proposal we employ is the one that identifies the complexity of a mixed state by the volume of codimensional-one hypersurface enclosed by Hubeny-Rangamani-Takayanagi surface. Using the finite difference method, the numerical calculations reveal that temperature, velocity, and subregion length all have an increasing effect on holographic subregion complexity. For arbitrary values of temperature and subregion length, as velocity approaches its relativistic upper limit, holographic subregion complexity exhibits a divergence. This divergence behavior observed in $d=2,3,4$ seems to demonstrate a universal behavior and is characterized by the Lorentz factor squared, $\gamma^{2}$.
\end{abstract}
\maketitle
%

{\textit{\textbf{Introduction}}}:
The AdS/CFT correspondence, holography, or more generally gauge/gravity duality serves as a valuable framework for exploring strongly coupled systems that are challenging to analyze using conventional perturbative methods. Gauge/gravity duality has been recognized as a strong-weak correspondence, linking a strongly coupled gauge theory in $d$-dimensional space-time with a classical gravity theory in $d+1$-dimensional space-time \cite{CasalderreySolana:2011us}. It has been effectively utilized to investigate various phenomena across fields, including condensed matter physics and low-energy quantum chromodynamics (QCD), where traditional perturbation techniques fall short. An illustrative example of such a system is the quark-gluon plasma (QGP) formed during heavy ion collisions which behaves as a strongly coupled fluid with low viscosity \cite{one}.

Interestingly enough, the holographic concept establishes an outstanding connection between quantities from quantum information theory and specific geometric measures in the gravity theory. Developing of this connection started by Hubney-Rangamani-Takayanagi (HRT) proposal \cite{taka1}. This proposal provides a simple geometric prescription of entanglement entropy, as a measure of quantum correlation of a pure quantum state, which has successfully withstood numerous tests \cite{mm1,mm2,mm3,mm4,mm5,mm6,mm7,mm8,mm9}. 
Complexity, as another key concept in quantum information theory, is a quantity that has also been studied holographically. Complexity is defined as the least number of simple gates necessary to create a given state from a reference state \cite{cc1}. This measure indicates the time and space resources required for efficient computation \cite{cc2}. In the context of quantum field theory, complexity refers to the minimum number of unitary operators needed to convert a reference state into a target state \cite{cc3}. Essentially, complexity can be used to categorize different quantum states according to the difficulty involved in their preparation. Within the holographic framework, two conjectures have been put forward to characterize complexity: the CV (complexity = volume) conjecture and the CA (complexity = action) conjecture. In the CA conjecture, complexity is determined by calculating the bulk action on the Wheeler-de-Witt patch, which is anchored at a specific boundary time. The CV conjecture defines complexity as the volume of a codimension-one hypersurface in the bulk that ends on a time slice of the boundary \cite{susskind1,susskind2}. Originally formulated for the complexity of pure states in the entire boundary system, both conjectures can be extended to include the complexity of mixed states in corresponding subregions \cite{comments,alishahiha}. Motivated by the HRT proposal, the CV proposal has been extended in such a way that the complexity of a subsystem at the boundary is defined by the volume of the codimension-one hypersurface  enclosed by the HRT surface \cite{alishahiha}, which we here refer to as holographic subregion complexity. Numerous studies examining the CV and CA conjectures, along with holographic complexity for mixed states across various gravity models, can be found in the literature \cite{v1,v2,v3,v4,v5,v6,v7,v8,v9,v10,v11,v12,v13,v14,v15}.

One of the initial motivations for studying the velocity dependence of quantities in holographic calculations was the fact that the quark-antiquark pair is not produced at rest in heavy ion collisions. At Super Proton Synchrotron (SPS) and Relativistic Heavy Ion Collider (RHIC) energy levels, the collective flow generated by the hot medium surrounding the quark-antiquark pair is considerable \cite{hotwind}. Based on this fact, studies were conducted, for example, to investigate the velocity dependence of the screening length of a moving heavy quark-anitquark pair \cite{hotwind}, as well as to examine the thermal width and the imaginary potential of a moving bound state in QGP \cite{AliAkabri}. On the other hand, the attention of holographic studies related to quantum information quantities turned to this field. In \cite{boost4}, to calculate the relative entropy between two states in the same Hilbert space, one of the states was considered to be a thermal moving plasma and the calculations at low temperatures and all orders in the velocity were performed. Other studies on such systems can be found in \cite{boost1, boost2}, where holographic entanglement entropy and holographic subregion complexity has been calculated in the perturbative regimes of small temperatures and velocities. Similar calculations in \cite{boost3} have been carried out for other mixed state information theoretical quantities such as the entanglement wedge cross section, mutual information, entanglement negativity, and purification complexity. In \cite{boost5}, the high temperature behavior of entanglement entropy and two point correlator at small velocities has been studied. A missed complete computation of holographic entanglement entropy, including arbitrary temperatures and velocities has been conducted in \cite{metric}. Inspired by this work, we have carried out a numerical computation for all temperatures and velocities to calculate holographic subregion complexity in a moving strongly coupled plasma, which appears to be absent in the literature.

Drawing inspiration from \cite{hotwind}, where high-velocity mesons are studied in a heavy ion collision and analyzed in the meson's rest frame which sees a moving hot wind, we similarly introduce a boosted black brane metric to provide a holographic description of a plasma moving with constant velocity. The aim we are pursuing is examining a mixed state within this plasma and address the question of how much information is needed to produce this state and how to quantify it using holographic complexity. The proposal we use to investigate the complexity of a mixed state is the proposal introduced in \cite{alishahiha}, which is detailed in holographic subregion complexity section. Thus, we start reviewing the background and then compute holographic subregion complexity and discuss its properties. The numerical method we used is described in Appendix \ref{A}.
\\


{\textit{\textbf{Review on the background}}}:\label{11}
In order to holographically describe a thermal strongly coupled plasma moving in a certain direction, characterized by the temperature $T$ and the velocity $v$, we study a $d+1$-dimensional boosted black brane as follows \cite{metric}
\begin{align}
ds^{2}=\frac{1}{z^2}\left(-dt^{2}+dx^{2}+g(z)(dt-v dx)^{2}+\frac{dz^{2}}{f(z)}+\sum_{i=1}^{d-2}dy_{i}^{2}\right),
\label{metric1}
\end{align}
where
\begin{align}
f(z)=1-\left(\frac{z}{z_{h}}\right)^{d},~~g(z)=\gamma^{2}\left(\frac{z}{z_{h}}\right)^{d},
\label{metric2}
\end{align}
in which, $z$ is the radial direction, $z_{h}$ is the position of the horizon, $\gamma=\frac{1}{\sqrt{1-v^2}}$ is the Lorentz factor and $0\leq v < 1$ is the boost parameter which is identified with the plasma velocity. The AdS boundary is approached as $z\rightarrow 0$ and the AdS radius is rescaled to be unit. According to gauge/gravity duality, the plasma lives on the $d$-dimensional boundary denoting by $(t,x,y_{1},...,y_{d-2})$ and the temperature of the black brane, corresponding to the temperature of the plasma is
\begin{align}
T=\frac{d}{4\pi z_{h}}.
\label{metric3}
\end{align}
Here we would like to consider the steady state case, where the plasma is moving with a constant velocity and the corresponding gravity background is not static but stationary, with the boost in the $x$-direction \eqref{metric1}. Clearly, in the case of $v\rightarrow 0$ the background reduces to the AdS black brane. On the other hand, in the case of $v \rightarrow 1$, the relativistic upper bound of velocity, $g(z)$ diverges due to the divergence of the Lorentz factor. This divergence can lead to numerical challenges in calculations of holographic subregion complexity at velocities approaching one, as we will see in the numerical result section.
 
This background has also been investigated to determine the relative entropy at low temperatures across all velocity orders \cite{boost4}, the behavior of holographic entanglement entropy and the two-point correlator at high temperatures and small velocities \cite{boost5}. In \cite{metric}, holographic calculations of entanglement entropy and mutual information has been done for all values of temperatures and velocities.\\

{\textit{\textbf{Holographic subregion complexity}}}:
It is important to highlight that in quantum systems, several methods have been developed to ensure that the concept of complexity for mixed states is independent of arbitrary degrees of freedom and can be reduced to the definition of complexity for pure states through a process known as purification. These methods include the spectrum approach, the purification approach, and the ensemble approach \cite{mixed}. Consequently, various interpretations of complexity for mixed states arise, and to our knowledge, a definitive explanation has yet to be established. In this paper, we describe complexity for mixed states as a criterion for measuring the amount of information needed to specify a particular mixed state. Essentially, it reflects the difficulty of generating and determining that mixed state. In light of these various interpretation, different proposals for defining complexity in the context of holography have been introduced \cite{comments}. Here, our focus is on a proposal inspired by the HRT proposal \cite{taka1}, which we refer to as holographic subregion complexity (HSC). In this proposal, the complexity of a subregion $A$ on the boundary is defined by the volume of codimensional-one hypersurface enclosed by HRT surface $\gamma_{A}$ that appears in the computation of holographic entanglement entropy \cite{alishahiha}, i.e. 
\begin{align}
\mathcal{C}_{A}=\frac{V_{\gamma_{A}}}{8\pi G_{N}},
\label{1b}
\end{align}
where $G_{N}$ and $\mathcal{C}_{A}$ are Newton constant and HSC for the subregion $A$, respectively.
\begin{figure}[H]
\centering
\includegraphics[width=55 mm]{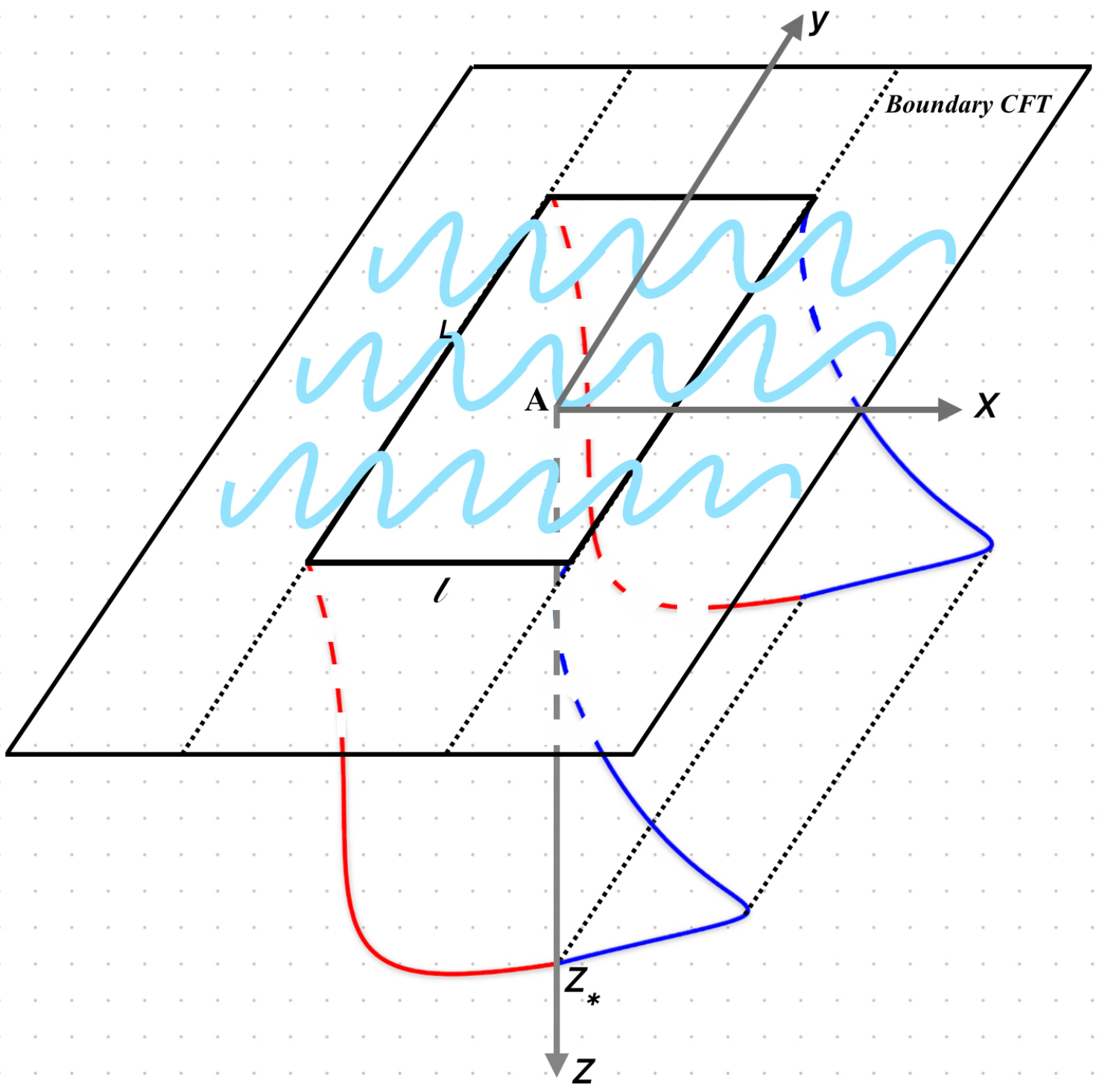}   
\caption{A schematic representation of the strip entangling surface of length $l$ and width $L\rightarrow \infty$ on a constant time slice of the AdS boundary. Our choice of subsystem is time-independent, while the underlying state is a moving plasma with constant velocity. The two solution branches have been shown by red and blue curves corresponding to plus and minus solutions, respectively.}
\label{fig1b}
\end{figure}
To study HSC, we would like to consider the subregion $A$ as a strip-like region in the field theory, i.e. on a constant time slice on the boundary of the boosted black brane \eqref{metric1}, as follows
\begin{align}
A:=x\in \left[-\frac{l}{2},\frac{l}{2}\right],~~~~y_{i}\in \left[-\frac{L}{2},\frac{L}{2}\right],~~~~i=1,...d-2,
\label{2b}
\end{align}
which is a strip-like surface of length $l$ and width $L\rightarrow \infty$ as shown schematically in figure \ref{fig1b}. Since the background \eqref{metric1} is not static, the extremal surface $\gamma_{A}$ moves out of a constant time slice in the bulk. Additionally, considering that for seeking simplicity the boost direction is taken along the $x$-direction which is the direction that the short edge of the strip is extended, there is a translational symmetry in the $y$-directions. Therefore, it is possible to represent $\gamma_{A}$ using the parametrization, $x=x(z)$ and $t=t(z)$. Using \eqref{metric1}, the area of the extremal surface is obtained
\begin{align}
S_{\gamma_{A}}=L^{d-2}\int \frac{dz}{z^{d-1}}\left(-(1-g)t'^{2}+(1+v^{2}g)x'^{2}-2vgt'x'+\frac{1}{f}\right)^{\frac{1}{2}},
\label{3b}
\end{align}
where $x'$ and $t'$ are derivatives with respect to $z$. The solutions of equation of motions, obtained by considering the integrand in equation \eqref{3b} as a Lagrangian, have two distinct branches $x_{\pm}(z)$ and $t_{\pm}(z)$ shown in figure \ref{fig1b} for which the boundary conditions for the two branches are enforced in the following manner
\begin{align}
x_{\pm}(z\rightarrow 0)=\pm \frac{l}{2},~~~~~t_{\pm}(z\rightarrow 0)=0,~~~~~x'_{\pm}(z\rightarrow z_{*})\rightarrow \infty,~~~~~t'_{\pm}(z\rightarrow z_{*})\rightarrow \infty,
\label{4b}
\end{align}
where $z_{*}$, turning point, is the point that two branches are smoothly joined. By numerically solving the equations of motion with boundary conditions \eqref{4b}, the profile of the extremal surface, $\gamma_{A}\equiv(x_{\pm}(z),t_{\pm}(z))$ is obtained \cite{metric}. Now, with the profile at hand, we parametrize the volume using $t=t(x,z)$, as we expected this function to be single-valued. In this case, the volume function used for finding the volume enclosed by $\gamma_{A}$ is
\begin{align}
V=\int dz dx \mathcal{V},
\label{5b}
\end{align}
where using \eqref{metric1},
\begin{align}
\mathcal{V}=\frac{1}{z^{2(d-1)}}\sqrt{(-(1-g)
   t_{x}^2-2 v g t_{x}+(1+v^2 g)) (-(1-g)
   t_{z}^2+\frac{1}{f})-t_{z}^2
   ((1-g) t_{x}+v g)^2},
\label{6b}
\end{align}
with $_{x}=\partial_{x}$ and $_{z}=\partial_{z}$. Then, the equation for the extremal solution is
\begin{align}
&z(f^2 ((v^2-1) g+1) t_{z}^3
   (4 (d-1) (v^2-1) g+(z-v^2 z)
   g_{z}+4 (d-1))+ f (t_{z} (g
    (4 (d-1) ((v^2-2) t_{x}^2+2 v
   t_{x}-2 v^2+1) \cr &
+ f (t_{z} (g
  (4 (d-1) ((v^2-2) t_{x}^2+2 v
   t_{x}-2 v^2+1)+(v^2-1) z g_{z}
   (v-t_{x})^2+4 z t_{xz}
   ((v^2-2) t_{x}+v)) \cr &
   - 4(v^2-1) g^2 (v-t_{x}) ((d-1)
   (v-t_{x})-z t_{xz})+4 (d-1)
   (t_{x}^2-1)+z g_{z} ((1-2
   v^2) t_{x}^2+2 v t_{x}+v^2-2) \cr &
   + 4 z t_{x} t_{xz})+2 z (g-1)
   ((v^2-1) g+1) t_{xx}
   t_{z}^2+2 z ((v^2-1) g+1)
   t_{zz} (g
   (v-t_{x})^2-t_{x}^2+1)) \cr &
   + z((v^2-1) g+1) (f_{z} t_{z}
  (g(v-t_{x})^2-t_{x}^2+1)+2
   t_{xx}))=0,
\label{1}
\end{align}
with the boundary condition determined by the HRT surface, i.e. $(x_{\pm}(z),t_{\pm}(z))$. We solve this equation numerically using the finite difference method, with details provided in the Appendix \ref{A} for $d=2$. Now, with the extremal solution in hand, we obtain the HSC using \eqref{5b} and \eqref{6b} and substituting the suitable value in \eqref{1b}. It is worth noting that, considering the fact that we are working with a stationary scenario, we expect the HSC to be independent of time. Moreover, since the HSC is divergent, it is convenient to examine the subtracted version of HSC, using \eqref{1b}, as follows
\begin{subequations}
\begin{align}
& \label{eq1} C\equiv\frac{8\pi G_{N}}{L}(\mathcal{C}-\mathcal{C}_{0})=\frac{1}{L}(V-V_{0}),\\
& \label{eq3} \hat{C}\equiv\frac{8\pi G_{N}}{L}(\mathcal{C}-\mathcal{C}_p)=\frac{1}{L}(V-V_{p}),
\end{align}
\end{subequations}
where $\mathcal{C}$, $\mathcal{C}_{0}$ and $\mathcal{C}_{p}$ are the HSC for $A$ in the boosted black brane \eqref{metric1}, non-boosted black brane metric and pure AdS background, respectively. The volumes are calculated with respect to the same boundary region such that $V$ in \eqref{5b} reduces to $V_{0}$ by setting $v$ equals to zero and reduces to $V_{p}$ by setting $v$ and $T$ equal to zero. $C$ and $\hat{C}$ represent the relative HSC of the mixed state in the plasma with constant temperature and velocity compared to that in a static plasma, and that in the zero-temperature static plasma, respectively. The reason for introducing two subtracted versions is to allow us to separately study the effects of velocity and temperature on the HSC.\\

{\textit{\textbf{Numerical results}}}:\label{2111}
In this section we present our results from a numerical calculation of HSC using the metric \eqref{metric1}, \eqref{eq1} and \eqref{eq3}. 

\begin{figure}[H]
\centering
\includegraphics[width=75 mm]{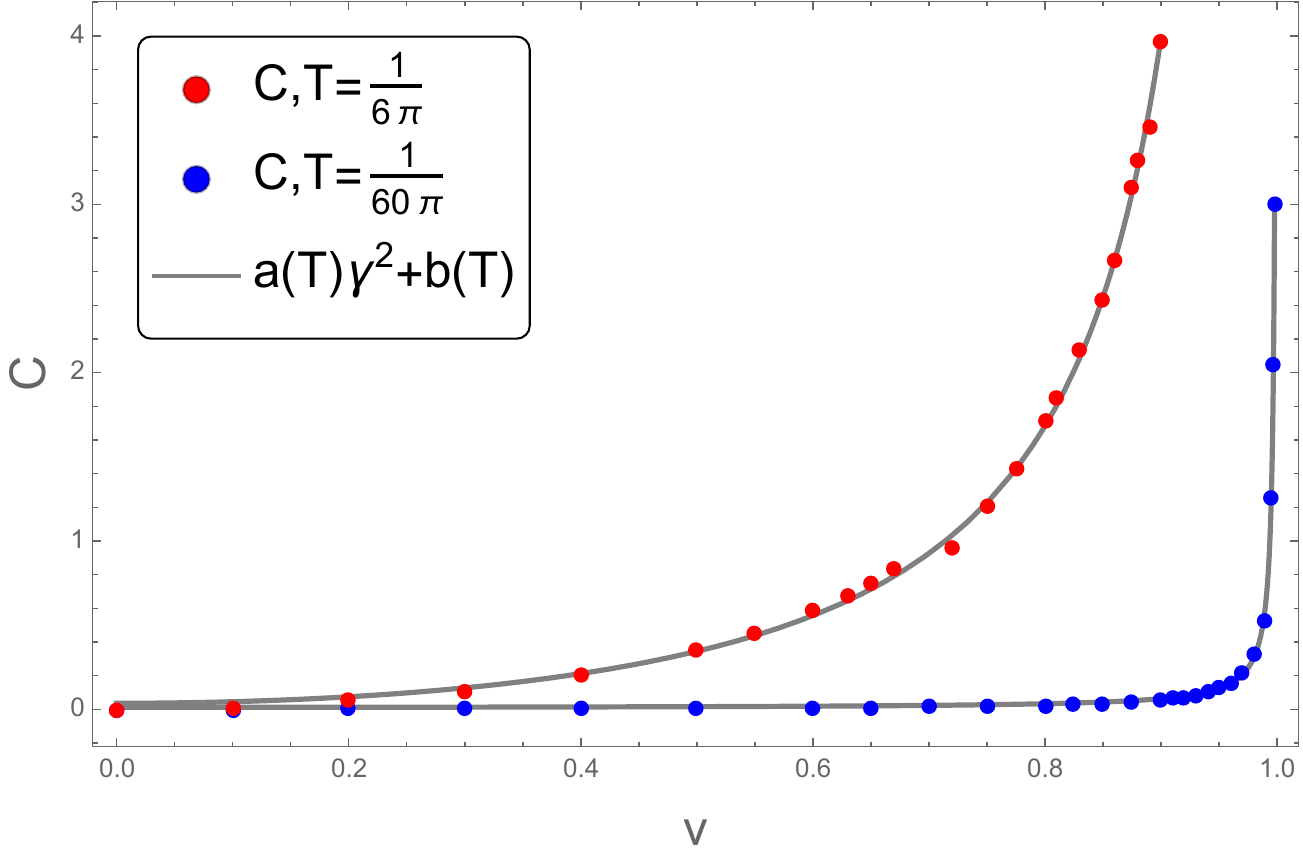} 
\includegraphics[width=75 mm]{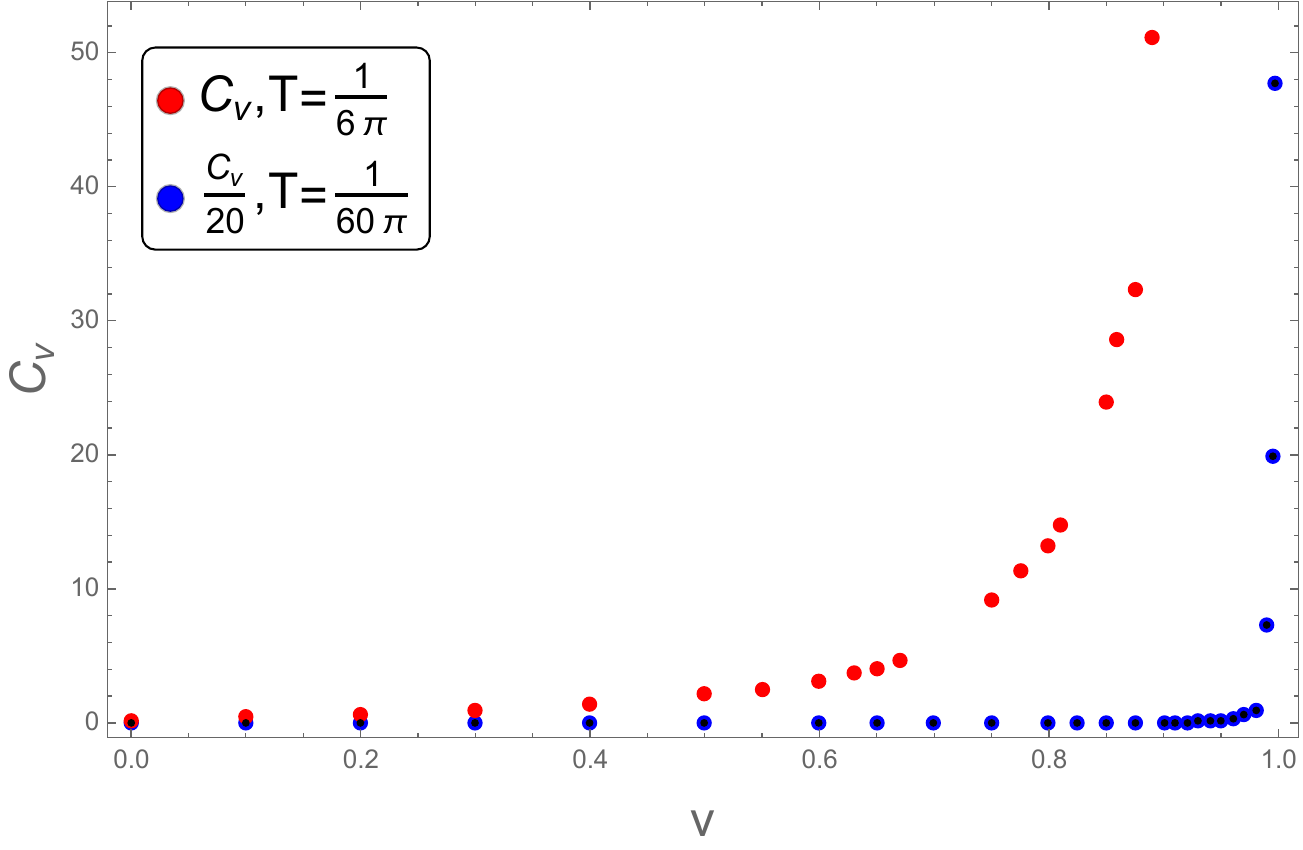}    
\caption{Left: The behavior of $C$ in terms of $v$ for two fixed values of $T$ and $l=1$. The red points and blue points represents numerical results for $C$ at $T=\frac{1}{6\pi}$ and $T=\frac{1}{60\pi}$, respectively. The gray curves are the function, $a(T)\gamma^{2}+b(T)$, fitted to the numerical data. For $T=\frac{1}{6\pi}$, $\mathcal{C}_{0}=157.5655$,
$a=0.9260$ and $b=-0.8898$ with a relative error, $|\frac{a\gamma^{2}+b-C(v)}{\mathcal{C}(v)}|$, whose mean value is $0.2344\%$ and maximum value is $0.4090\%$. For $T=\frac{1}{60\pi}$, $\mathcal{C}_{0}=209.5485$, $a=0.0121$ and $b=-0.00884$ with a relative error, whose mean value is $0.0070\%$ and maximum value is $0.0403\%$. Right: The slope of $C$, $C_{v}\equiv dC/dv$, as a function of $v$ for $l=1$ and two fixed values of temperatures.}
\label{fig1}
\end{figure}

{\textit{\textbf{d=2 moving plasma}}}:
In the left panel of figure \ref{fig1}, we have plotted $C$ as a function of the plasma velocity $v$ for two fixed values of temperatures. As the figure shows, $C$ is positive, which according to \eqref{eq1}, means that the complexity of the mixed state in the plasma with a constant velocity is always greater than the complexity of the same state in the same plasma with zero velocity. Therefore, we need more information to specify a mixed state in the presence of velocity. Moreover, with rising $v$ the amount of information it takes to specify the mixed state increases. Additionally, with the increase in $v$, the dependence of $C$ on temperature becomes more noticeable in such a way that, at a sufficiently large value of $v$, the amount of information required to determine the mixed state is significantly greater at higher temperatures compared to a plasma with a lower temperature. Another point that can be observed from this figure is that as $v$ approaches one, $C$ increases with respect to $v$ dramatically.

We have fitted the numerical values obtained for $C$ in terms of $v$ with a function of the form $a(T)\gamma^{2}+b(T)$ for two temperatures, as shown in the left panel of figure \ref{fig1}. Based on numerical results, we observe that $a$ and $b$ are of the same order and $a,b \ll \mathcal{C}_{0}$. Then, using \eqref{eq1}, $\mathcal{C}-\mathcal{C}_{0}\propto a(T)\gamma^{2}+b(T)$ can effectively be considered in the form of $C\propto a(T)\gamma^{2}$. In fact, our claim is that the presence of $b$, since it is very small compared to $\mathcal{C}_{0}$, is due to the error in the precision of our numerical calculations. According to this fitted function, $C$ diverges as the velocity approaches its relativistic upper bound $v\rightarrow 1$ or the Lorentz factor $\gamma\rightarrow \infty$, as it is expected and in agreement with the left panel of figure \ref{fig1}.\footnote{Due to numerical challenges, unfortunately it is not possible to calculate $C$ for velocities close to one at higher temperatures. Because in the limit of $v\rightarrow 1$, the metric coefficient $g(z)$ diverges. It seems that this divergence has a chance to be suppressed at low temperatures but it is not suppressed at high temperatures in the process of calculating complexity and leads to numerical challenges.} As shown in the right panel of figure \ref{fig1}, the slope of the points for these two temperatures changes very rapidly when $v\rightarrow 1$. This divergence is characterized by $\gamma^{2}$ based on our numerical results. 
The $\gamma^{2}$ behavior is also reported for the subtracted version of holographic entanglement entropy, $\Delta S\equiv S-S_{p}$, at small temperatures and large velocity limit, \cite{metric}. $S$ is the entanglement entropy for a strip-like boundary subregion aligned with the boost direction in the boosted black brane geometry and $S_{p}$ is the entanglement entropy for the same boundary region in pure $\rm{AdS}$ background. However, using an analytical calculation in $d=2$, it has been shown that the leading order divergence in $\Delta S$ varies as $\gamma$ for large velocity at arbitrary temperatures. As mentioned, we cannot probe the velocity close to one for any desired temperature due to numerical challenges. Thus, we do not add any remarks about the change in divergence behavior of HSC with respect to $\gamma$ at higher temperatures.

Another point is the result we want to report from \cite{boost3}, where HSC in a moving plasma has computed perturbatively as the leading order change over pure AdS. In this reference, the direction of boost in the boosted black brane geometry is compactified on a circle. The subtracted version of HSC, $\Delta C\equiv C-C_{p}$, where $C$ is the HSC for a strip-like boundary subregion along the boost direction in the boosted black brane background and $C_{p}$ is the HSC for the same boundary region in pure AdS geometry, is analytically calculated. Interestingly, the obtained analytical function for $\Delta C$ is proportional to $\gamma^{2}$, where the constant term of the function becomes zero for $d=2$ similar to the result we obtained in our model.

\begin{figure}[H]
\centering
\includegraphics[width=75 mm]{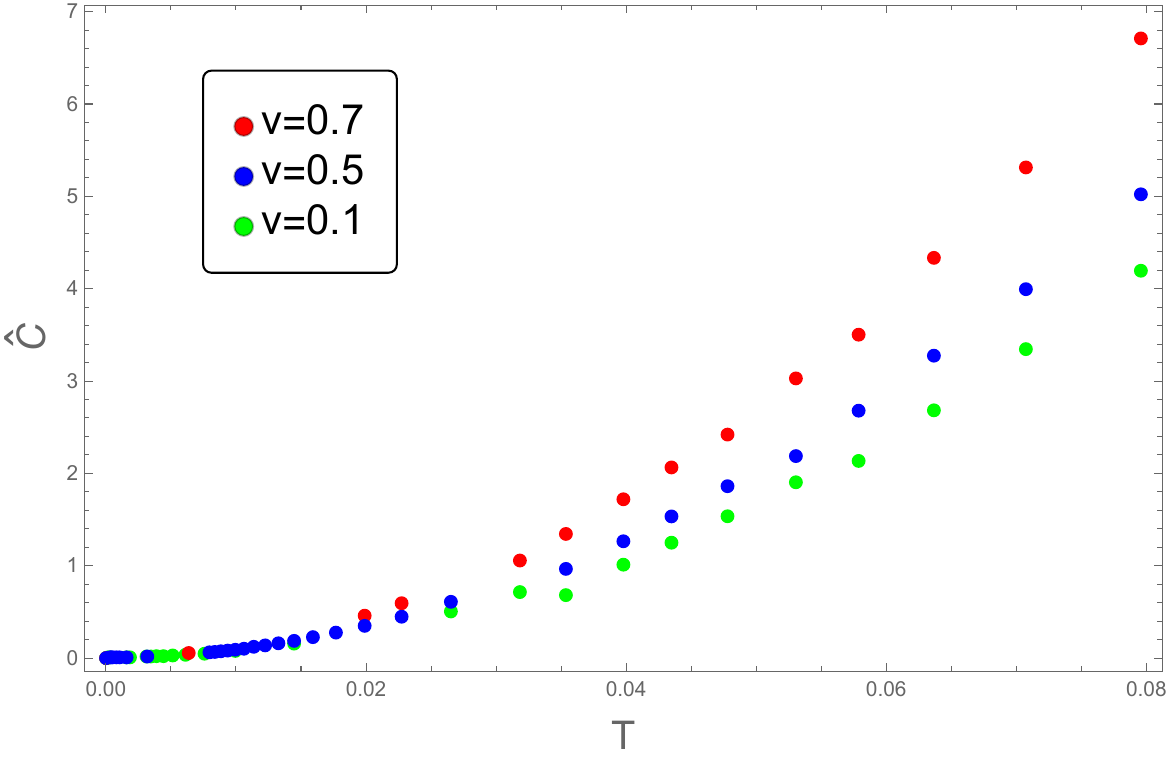}   
\caption{The behavior of $\hat{C}$ as a function of $T$ for $l=1$ and three fixed values of velocities.}
\label{fig2}
\end{figure}
In figure \ref{fig2}, we have presented a plot of the behavior of $\hat{C}$, \eqref{eq3}, in terms of temperature. It is illustrated by this figure that with the increase in temperature, $\hat{C}$ increases. It can also be observed that for a given value of temperature, the higher the plasma velocity, the more the information needed to determine the mixed state in agreement with previous results. Moreover, according to this figure, at low temperatures the dependence of $\hat{C}$ on velocity is suppressed. This result can also be analyzed and supported analytically. Within the limit of $T\rightarrow 0$ (or $z_{h}\rightarrow \infty$), the metric coefficients $f(z)\rightarrow 1$ and $g(z)\rightarrow 0$, as evident from \eqref{metric2}. In this limit, the equation for the exteremal surface \eqref{1} takes the following form for $v<1$,
\begin{align}
2t_{z}^{3}+2t_{z}(t_{x}^{2}+zt_{x}t_{xz}-1)-zt_{z}^{2}t_{xx}+z(t_{xx}-t_{zz}(t_{x}^{2}-1))=0.
\label{N1}
\end{align}
Clearly, in this limit the equation of motion \eqref{N1} is independent of both $T$ and $v$. The suppression of $v$ occurs because the product of $g(z)$ and $v$ always appears in \eqref{1} and $g(z)$ goes to zero. As a result, although $\hat{C}$ has been plotted in terms of temperature, its value is zero at zero temperature.

\begin{figure}[H]
\centering
\includegraphics[width=75 mm]{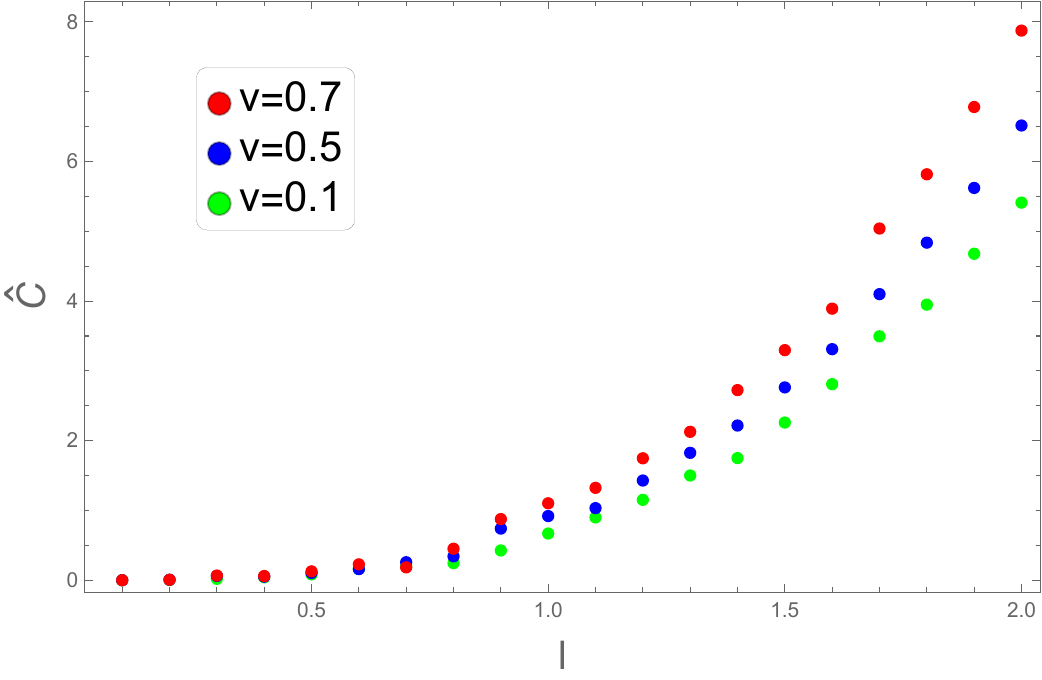} 
\includegraphics[width=75 mm]{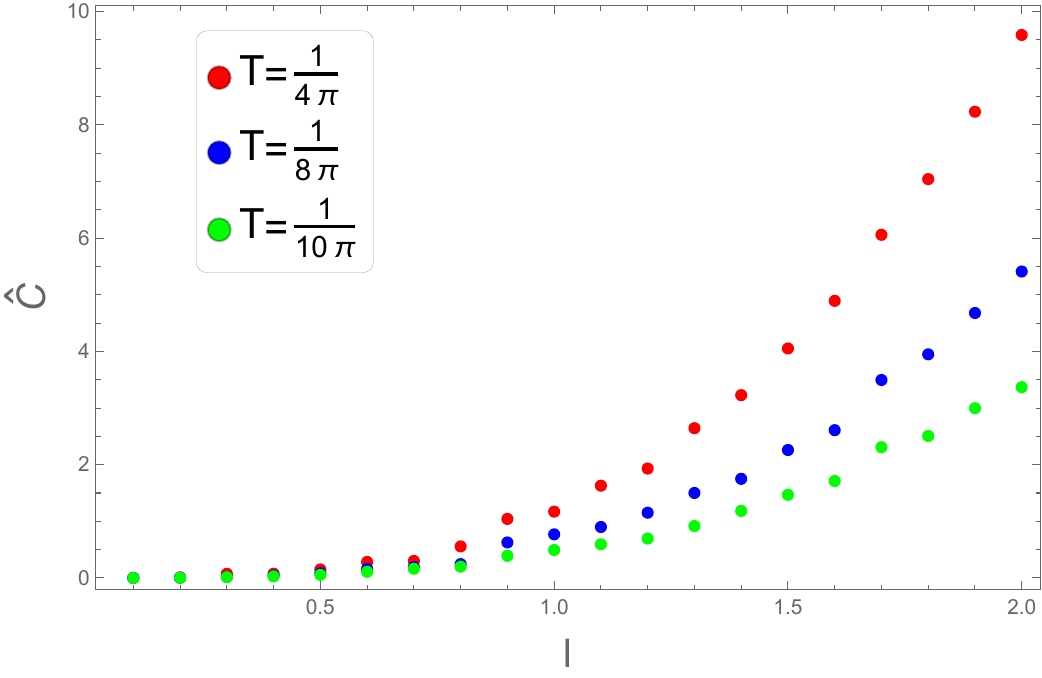}    
\caption{The behavior of $\hat{C}$ in terms of $l$ for different values of $v$ at $T=\frac{1}{8\pi}$ (left) and for different values of $T$ at $v=0.1$ (right).}
\label{fig3}
\end{figure}

To examine the effect of subregion length $l$ on HSC in figure \ref{fig3}, $\hat{C}$ is plotted as a function of $l$. As shown in these two panels, $\hat{C}$ grows as $l$ becomes larger. In other words, the information required to determine the mixed state in the plasma with fixed temperature across various velocities (left panel) and in the plasma with fixed velocity across different temperatures (right panel) increases as $l$ grows. According to the left (right) panel of figure \ref{fig3}, for given $l$ and fixed temperature (velocity), the information required to specify the mixed state increases with rising velocity (temperature) of the plasma, in agreement with the left panel of figure \ref{fig1} and figure \ref{fig2}.

In figure \ref{fig3}, it is shown that whether at a given value of temperature for various velocities (left panel) or for a fixed value of velocity for different temperatures (right panel), $\hat{C}$ approaches zero for a sufficiently small $l$. On the field theory side, the high energy limit of the theory is probed by small values of $l$. In this limit, the energy scale which we assign to $l$, called $T_{l}\equiv l^{-1}$, should be very larger than the temperature of the plasma, $T_{l}\gg T$. Therefore, in this limit the impact of $T$ is effectively suppressed on HSC. To understand this point, we can consider the metric coefficients in this limit, as $T$ approaches zero. According to \eqref{metric2}, $f(z)$ and $g(z)$ goes to one and zero, respectively which leads to a reduction in the effects of temperature and also velocity on HSC. On the gravity side, in this limit $z_{*}$ is close to the boundary. Consequently, the vicinity of the boundary region, which closely resembles pure AdS is realized by the extremal surface or equivalently HSC and as a result $\mathcal{C} \rightarrow \mathcal{C}_{p}$ in \eqref{eq3}. On the other hand, the larger $l$ leads to significant effects of $T$ and $v$ on HSC because of the probing of the low energy limit of the theory, i.e. $T_{l}\approx T$. On the gravity side, in this limit $z_{*}$ extends deeper into the bulk, allowing for more deviation from pure AdS geometry that can be realized more by the extremal surface, as is observed in two panels of figure \ref{fig3} for larger values of $l$.

\begin{figure}[H]
\centering
\includegraphics[width=75 mm]{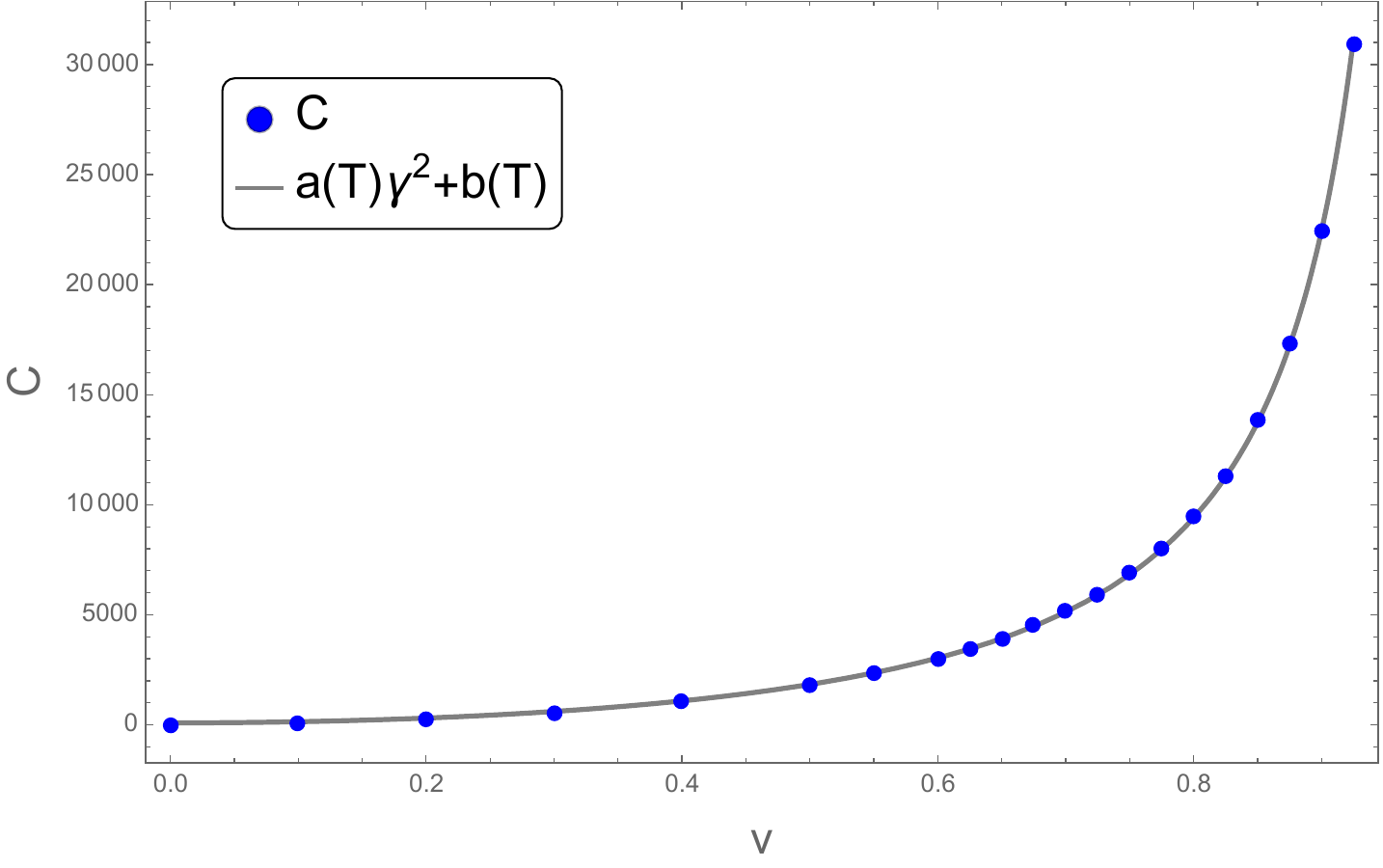} 
\includegraphics[width=75 mm]{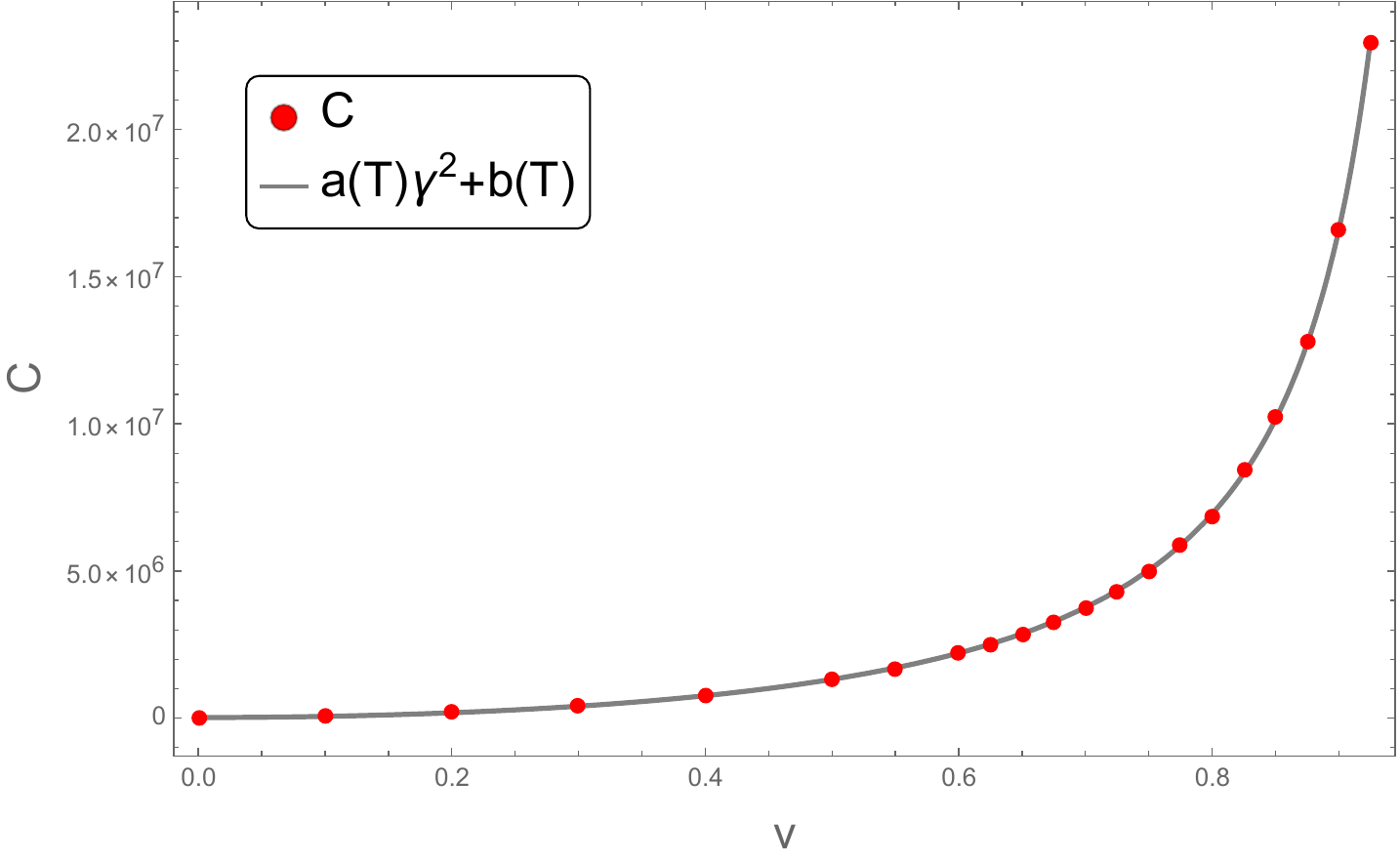}   
\caption{Left: $C$ in terms of $v$ at $T=\frac{1}{4\pi}$, $l=1$ and $d=3$. The blue points are numerical results and $\mathcal{C}_{0}=677549.3917$. The gray curve is the function, $a(T)\gamma^{2}+b(T)$, fitted with the data by $a=5239.6824$ and $b=-5150.1775$ with a relative error whose mean value is $0.0088\%$ and maximum value is $0.0318\%$. Right: $C$ in terms of $v$ at $T=\frac{1}{4\pi}$, $l=1$ and $d=4$. The red points are numerical results and $\mathcal{C}_{0}=1.1273\times 10^{9}$. The gray curve is the function, $a(T)\gamma^{2}+b(T)$, fitted with the data by $a=3.8888\times 10^{6}$ and $b=-3.8663\times 10^{6}$ with a relative error whose mean value is $0.0032\%$ and maximum value is $0.0139\%$.}
\label{fig4}
\end{figure}

{\textit{\textbf{d=3,4 moving plasma}}}:
In this section, we have focused on repeating the numerical computations in three and four dimensions. In figure \ref{fig4}, we have plotted $C$ in terms of $v$ for $d=3$ (left panel) and $d=4$ (right panel). The result is similar to the outcome obtained in case $d=2$, with the exception that the absolute value of HSC increases significantly as the dimension increases. As this figure shows, $C$ grows with $v$ and this increase becomes very steep at velocities close to one. We have fitted the numerical values of $C$ in terms of $v$ using a function of the form $a(T)\gamma^{2}+b(T)$. We notice that, similar to $d=2$, $C$ diverges as $v\rightarrow 1$. Our analysis suggests that this divergence is proportional to $\gamma^{2}$. This may represent a universal behavior for the strip-shaped subsystems we are examining with the plasma moving parallel to its short edge. Exploring this divergence in relation to more general subsystem geometries would be interesting.

\begin{figure}[H]
\centering
\includegraphics[width=75 mm]{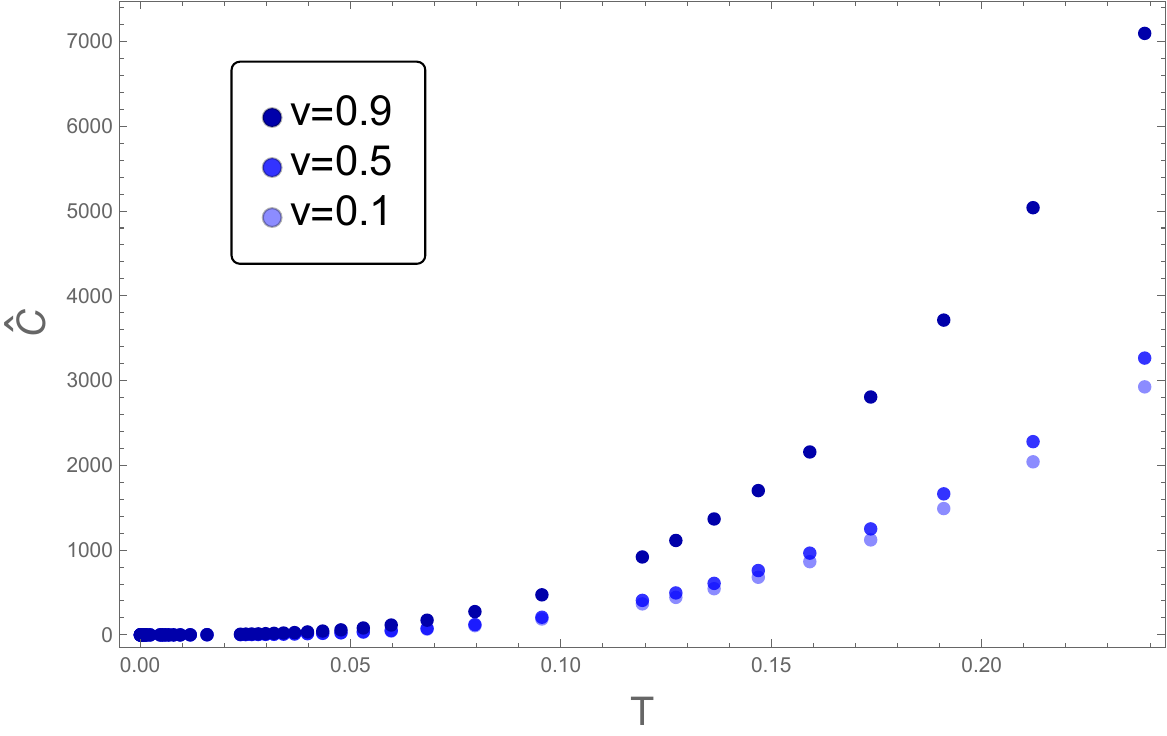} 
\includegraphics[width=75 mm]{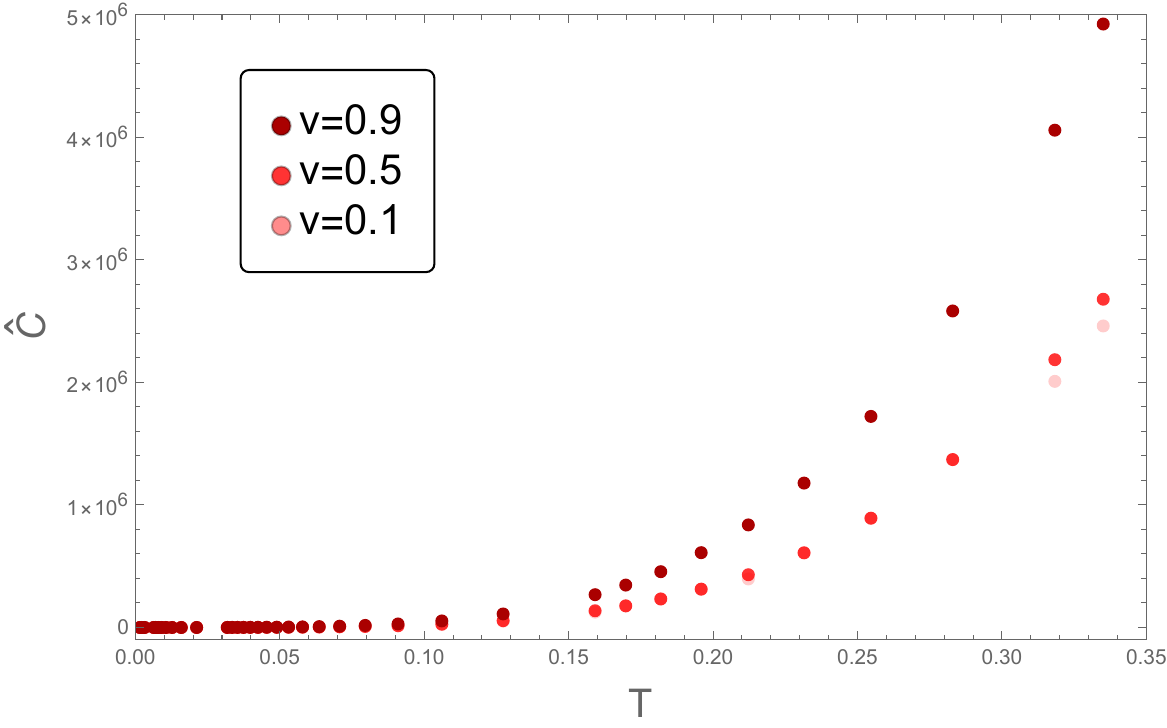}   
\caption{The behavior of $\hat{C}$ in terms of $T$ for $l=1$ and three fixed values of $v$ at $d=3$ (left) and $d=4$ (right).}
\label{fig5}
\end{figure}

In figure \ref{fig5}, $\hat{C}$ is plotted as a function of $T$ at $d=3$ (left panel) and $d=4$ (right panel). The obtained results show that the absolute values of HSC have increased significantly at $d=3,4$ and similar to the case of $d=2$ temperature has an increasing effect on HSC. By comparing these panels with figure \ref{fig2}, it can be observed that as the dimension increases, the distinction between the values of $\hat{C}$ at different velocities becomes more pronounced at higher temperatures. It can also be seen that this difference in the values of $\hat{C}$ is less noticeable at lower velocities.

\begin{figure}[H]
\centering
\includegraphics[width=75 mm]{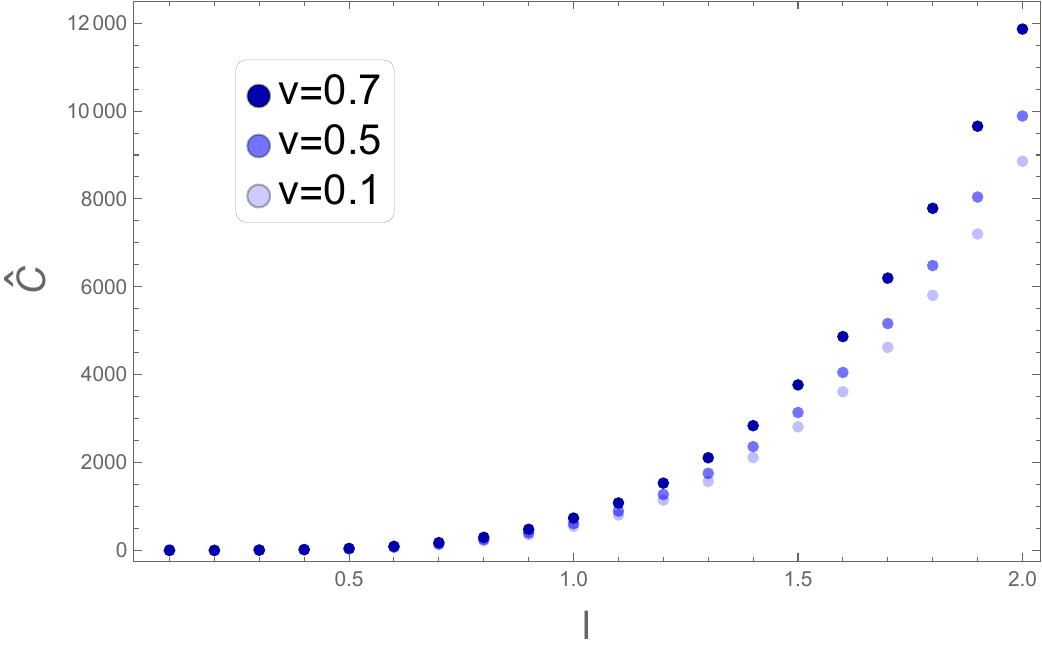} 
\includegraphics[width=75 mm]{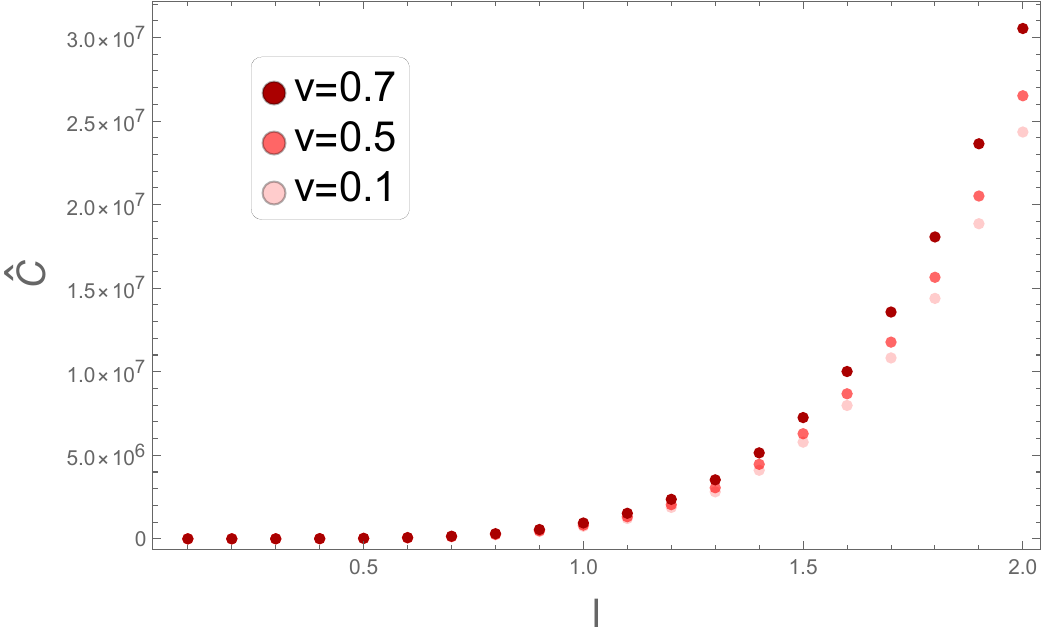}   
\caption{The behavior of $\hat{C}$ as a function of $l$ for different values of $v$ at $T=\frac{3}{20\pi}$ and $d=3$ (left), at $T=\frac{1}{4\pi}$ and $d=4$ (right).}
\label{fig6}
\end{figure}

\begin{figure}[H]
\centering
\includegraphics[width=75 mm]{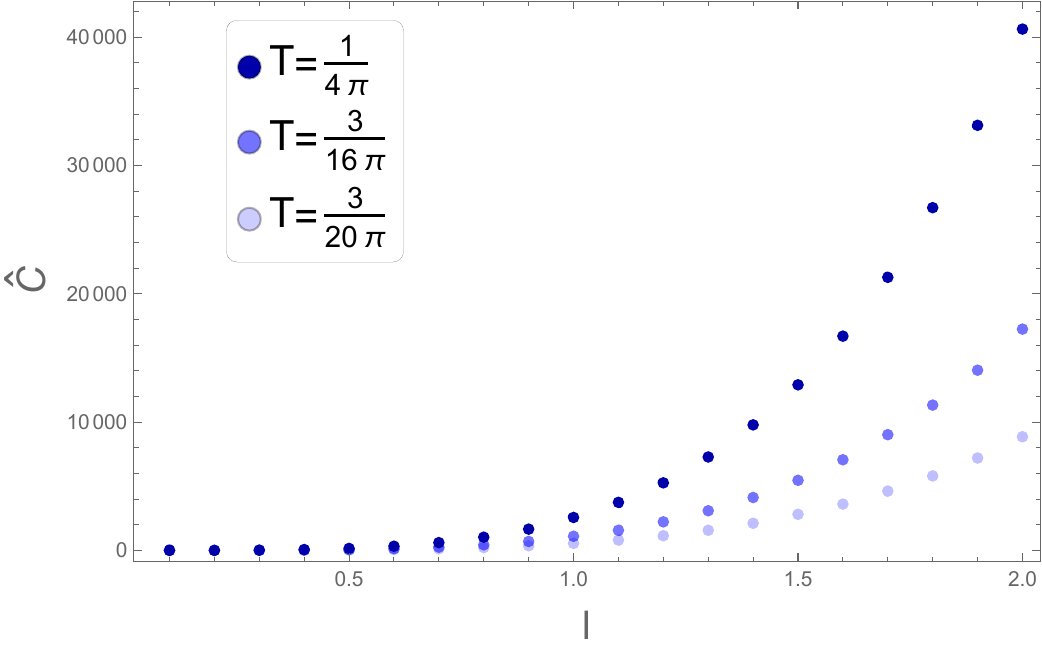} 
\includegraphics[width=75 mm]{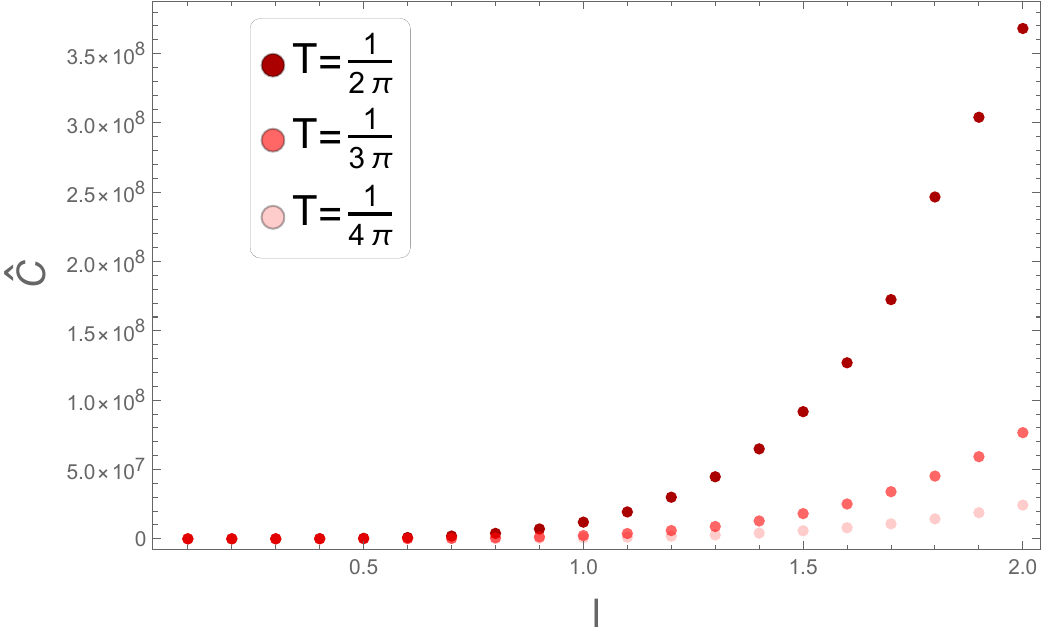}   
\caption{The behavior of $\hat{C}$ as a function of $l$ for different values of $T$ at $v=0.1$ for $d=3$ (left), and $d=4$ (right).}
\label{fig7}
\end{figure}

In figures \ref{fig6} and \ref{fig7}, we have presented a plot of the behavior of $\hat{C}$ as a function of subregion length $l$ for $d=3$ (left panels) and $d=4$ (right panels). In figure \ref{fig6}, this function is plotted for three different values of velocities at a fixed value of temperature, whereas in figure \ref{fig7}, it is depicted for three different values of temperatures at a fixed velocity. In comparison with figure \ref{fig3}, it becomes apparent that as the dimension increases, the difference between the values of $\hat{C}$ at different velocities (figure \ref{fig6}) and different temperatures (figure \ref{fig7}) becomes more noticeable with larger values of $l$. Furthermore, as the dimension increases, the distinction between the values of $\hat{C}$ for different velocities becomes smaller. Other results are similar to those related to the $d=2$ case.

{\textit{\textbf{Conclusions}}}:
In this paper, we investigate holographic subregion complexity in a moving strongly coupled plasma across dimensions $d=$ 2, 3, and 4. Using the proposal introduced in \cite{alishahiha} for holographic subregion complexity, we considered a strip-like subsystem in such a way that the plasma moves at a constant velocity parallel to its short edge. Previous studies of systems like this have relied on certain simplifying assumptions regarding the parameters of the plasma \cite{boost2,boost3}. Not considering these limitations on the parameters in calculating holographic subregion complexity is a novel aspect of our research. Our numerical calculations demonstrate that temperature, velocity, and subregion length all contribute to an increase in holographic subregion complexity. Additionally, as the velocity of plasma approaches its relativistic upper limit, holographic subregion complexity exhibits a divergence that is proportional to the Lorentz factor squared, $\gamma^{2}$. This divergence appears to exhibit universal behavior in dimensions $d=$ 2, 3, and 4, as well as a universal behavior across all temperatures and subregion lengths in each dimension. Among the interesting studies that can be conducted in the future in this context is the examination of holographic subregion complexity for subregions with arbitrary geometric shapes, for more complex plasmas beyond steady-state flows, and for cases where the plasma velocity has an arbitrary angle with respect to the subregion of interest.

\appendix
\section{Numerical method}\label{A}

In this appendix, we explain our numerical method to solve the equation \eqref{1} for the extremal solution. The HRT surface, $(x_{\pm}(z),t_{\pm}(z))$ has been calculated in \cite{metric}. In order to obtain the HSC, as the volume enclosed by the HRT surface, our computation resulted in a complicated nonlinear partial differential equation \eqref{1}, that is not solvable by conventional methods. The method we used is the finite difference method (FDM) implemented in Mathematica. The finite difference techniques are based on approximations in which differential equations are replaced with finite difference equations. FDM generally consists of three main steps. The first step includes segmenting the solution area into a grid of nodes. The second step involves the approximation of the specified differential equation using a finite difference formulation that connects the dependent variable at a specific point in the solution area to its values at adjacent points. The third or final step is solving the resulting difference equations while adhering to the given boundary conditions and/or initial conditions \cite{A1}.

\begin{figure}[H]
\centering
\includegraphics[width=60 mm]{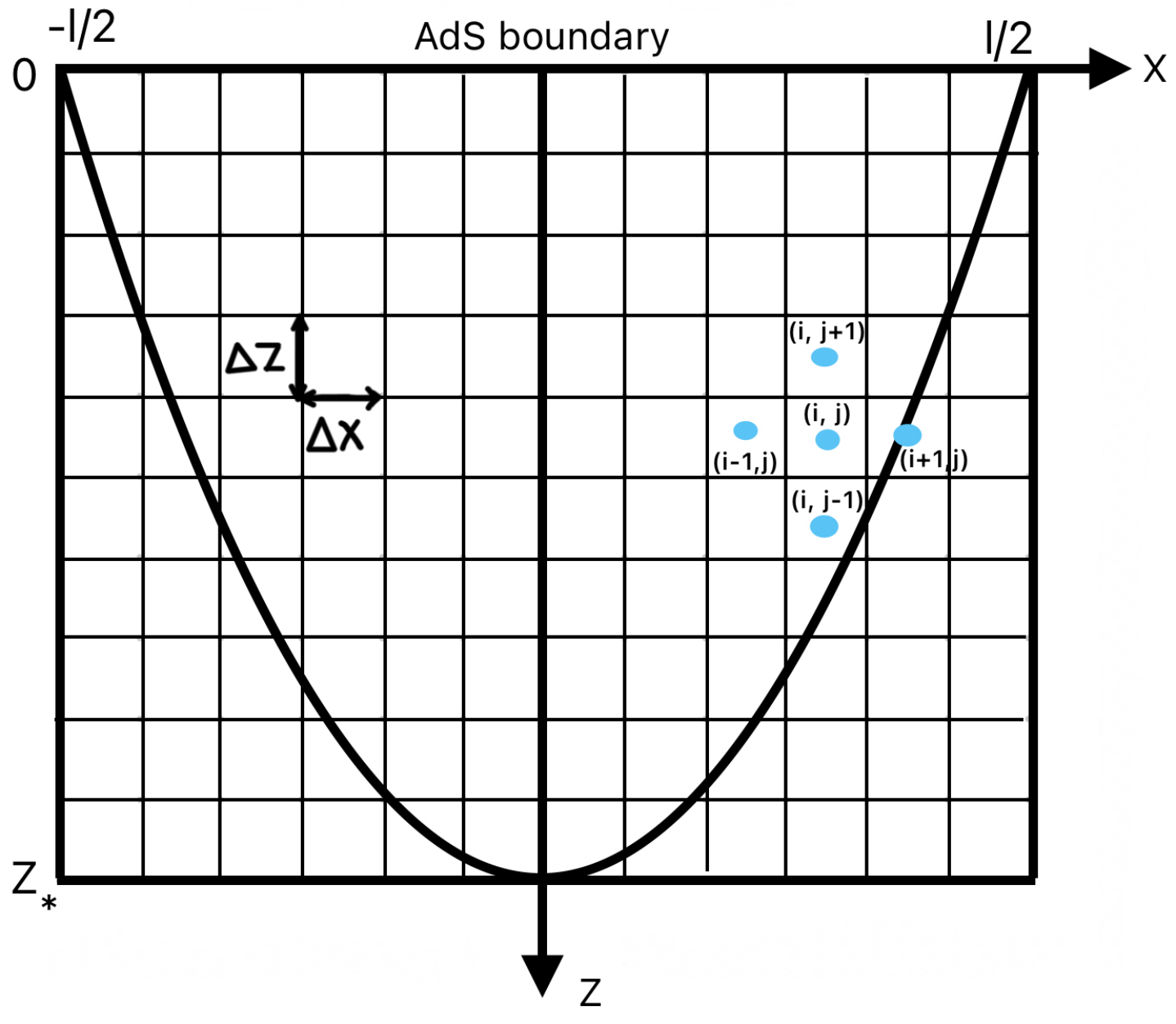}
\caption{A schematic representation of the finite difference mesh of the solution area. The curve schematically represents $x(z)$ as the boundary condition and the $AdS$ boundary located at $z=0$. The nodes located in the cells through which the boundary curve passes are used as fixed nodes to determine the unknown ones.}
\label{fig1aa}
\end{figure}

Unlike the conventional approaches used in FDM, the boundary condition of our equation is determined by HRT surface. Therefore, our first challenge in applying this method to solve the equation and find $t(x,z)$ is defining and discretizing the boundary condition. In the first step we divide the solution area into grid nodes:
\begin{align}
& x=i\Delta x,~~~~~~~~~~~~~i=0,1,2,...,n_{x},\nonumber\\
& z=j\Delta z,~~~~~~~~~~~~~j=0,1,2,...,n_{z},
\label{aa1}
\end{align}
where $\Delta x$ and $\Delta z$ are the mesh sizes in the $x$ and $z$ directions and $t(i\Delta x,j\Delta z)\equiv t(i,j)$. This meshing is designed in such a way as to ensure that the grid nodes pass through the boundary curve. For this purpose, by selecting an arbitrary value for $n_{z}$, the value of $n_{x}$ is determined using $x_\pm(z)$ and based on the accuracy governed by the cell sizes the set of $\{i,j,t(i,j)\} $ is constructed as the discrete boundary. Just as is evident from the boundary condition, the range of variation for $x$ and $z$ are $[-\frac{l}{2},\frac{l}{2}]$ and $[0,z_{*}]$, respectively and our discretization included cells of maximum size $\mathcal{O}(10^{-3})$. Therefore, as schematically shown in figure \ref{fig1aa}, by discretizing the boundary condition we use these nodes as known or fixed nodes to find the unknown nodes within the solution area. 

In the second step, we approximate our target differntial equation, i.e. \eqref{1} with a finite difference form. To do so, we use the central difference approximation of the derivatives of $t$ at the $(i,j)$th node which are
\begin{align}
& t_{x}(i,j)\approx \frac{t(i+1,j)-t(i-1,j)}{2\Delta x}+\mathcal{O}(\Delta x)^{2},\nonumber\\
& t_{z}(i,j)\approx \frac{t(i,j+1)-t(i,j-1)}{2\Delta z}+\mathcal{O}(\Delta z)^{2},\nonumber\\
& t_{xx}(i,j)\approx \frac{t(i+1,j)-2t(i,j)+t(i-1,j)}{(\Delta x)^{2}}+\mathcal{O}(\Delta x)^{2},   \nonumber\\
& t_{zz}(i,j)\approx \frac{t(i,j+1)-2t(i,j)+t(i,j-1)}{(\Delta z)^{2}}+\mathcal{O}(\Delta z)^{2},    \nonumber\\
 &t_{xz}(i,j)\approx \frac{t(i+1,j+1)-t(i+1,j)-t(i,j+1)+2t(i,j)-t(i-1,j)-t(i,j-1)+t(i-1,j-1)}{2\Delta x \Delta z}
\cr &~~~~~~~~~~ +\mathcal{O}[(\Delta x)^{2},(\Delta z)^{2}],
\label{a1}
\end{align}
where the indices of $t(i,j)$, corresponding to $x$ and $z$, represent the first derivative with respect to $x$ and $z$ and two indices indicate the second derivative. Substituting the expressions of \eqref{a1} into \eqref{1}, we found a discrete nonlinear equation for $t(i,j)$.

\begin{figure}[H]
\centering
\includegraphics[width=80 mm]{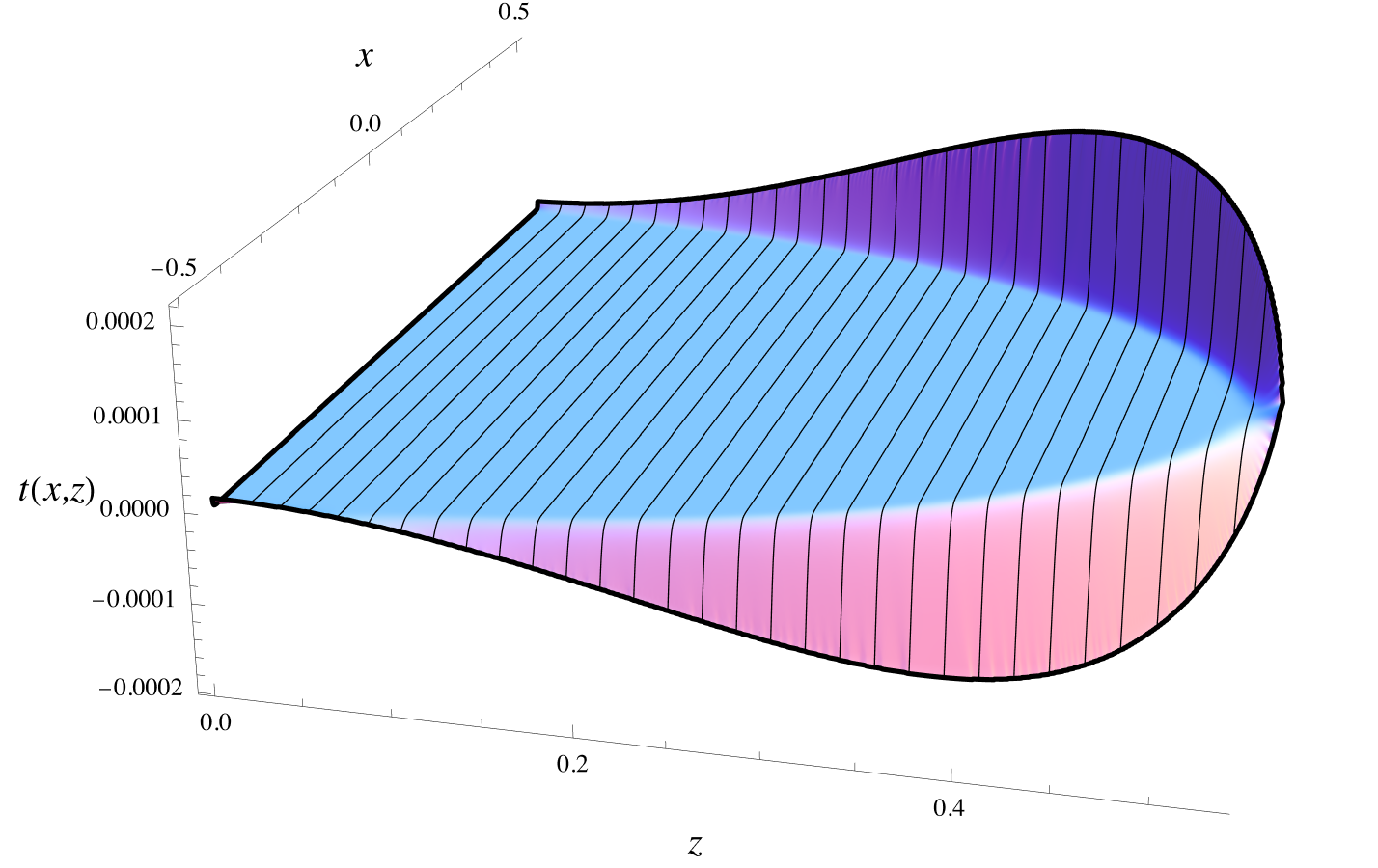}  
\includegraphics[width=80 mm]{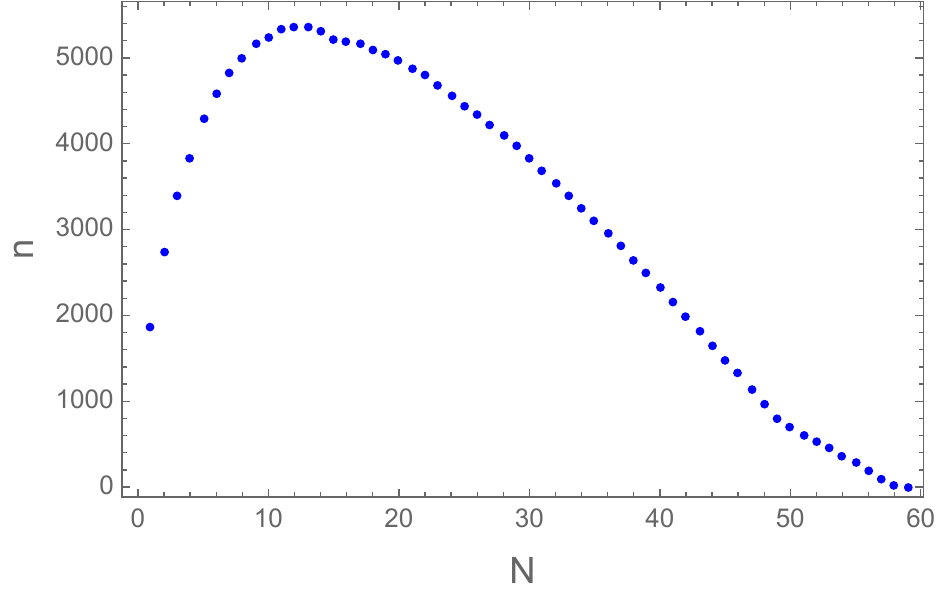} 
\caption{Left: An example solution in $d=2$, which is generated with $n_{x}=744$, $n_{z}=400$ and the number of iterations $N=60$ for $T=\frac{1}{6\pi}$ and $l=1$, when the fluid is moving with $v=0.1$. Maximum size of cell is $\mathcal{O}(10^{-3})$ and the tolerance is $\mathcal{O}(10^{-6})$. Right: The number of unrelaxed grid nodes in the solution area, $n$ in terms of the number of iterations, $N$ for the example solution displayed in the left panel.}
\label{fig1a}
\end{figure}

In the final step in order to solve the obtained discrete nonlinear equation, we used an iteration method. For $N$ iterations, the equation is solved $N$ times for all grid nodes in the solution area. After each iteration, the value of each node obtained from the previous iteration is compared with the value of the same node in the most recent iteration. As the convergence value approaches the correct value this difference tends to approach zero and in our numerical computation we used a tolerance of $\mathcal{O}(10^{-6})$. An example solution is illustrated in the left panel of figure \ref{fig1a}. In the right panel of figure \ref{fig1a}, the behavior of  the number of unrelaxed grid nodes in the solution area, $n$ as a function of $N$ for the example solution displayed in the left panel, has been plotted. As shown in this figure, considering that initially all grid nodes are assumed to be zero, $n$ increases with the rise of $N$. However, after approximately ten iterations, $n$ starts to decrease. Eventually, when $N$ reaches sixty, $n$ drops to zero and with the relaxation of all grid nodes, the solution is obtained.

Achieving the final relaxed system resulting from the iterations presents numerical challenges in the values of $lT$ for velocities close to one and a mesh size of $10^{-3}$. In other words, at near one velocities, this challenge arises in the high temperature range across all dimensions. However, in $d=2$ the maximum temperature attainable for probing velocities near one is lower than in $d=3,4$. As a result, higher dimensions allow for a broader and more accurate exploration of velocities approaching one. This is clearly illustrated by comparing figures \ref{fig1} and \ref{fig4} in the main text. Another point we would like to highlight is that, in order to keep the accuracy of the calculations or the cell size fixed, which in our work is chosen to be of the order of $10^{-3}$, the appropriate number of mesh points must be adjusted for each subregion length when calculating the HSC in terms of the subregion length.

A method similar to what we have done here can be found in \cite{vad}, where the time-dependent extremal volume surface is obtained by numerically solving a partial differential equation with boundary condition given by the HRT surface. They discretized the solution area, as an adaptive triangulation of the HRT surface, and solved the equation using the finite-element method in the $\rm{AdS_{3}}$ Vaidya geometry.

\begin{thebibliography}{99}

\bibitem{CasalderreySolana:2011us}
  J.~Casalderrey-Solana, H.~Liu, D.~Mateos, K.~Rajagopal and U.~A.~Wiedemann,
  ``Gauge/String Duality, Hot QCD and Heavy Ion Collisions,'' Cambridge University Press, 2014,
   \href{http://arxiv.org/abs/1101.0618}{[arXiv:1101.0618 [hep-th]]}.

\bibitem{one}
  R.~Baier, A.~H.~Mueller, D.~Schiff and D.~T. Son, 
  ``Bottom up thermalization in heavy ion collisions,'' Phys.~Lett.~B {\bf 502}, 51 (2001)
   \href{http://arxiv.org/abs/0009237}{[arXiv:0009237 [hep-th]]}.


\bibitem{taka1}
  Veronika E. Hubney, Mukund Rangamani, Tadashi Takayanagi,
  ``A Covariant holographic entanglement entropy proposal,'' 
  JHEP {\bf 07} (2007) 062,
   \href{http://arxiv.org/abs/0705.0016}{[arXiv:0705.0016[hep-th]]}.
  
\bibitem{mm1} 
  Mukund Rangamani and Tadashi Takayanagi,
  ``Holographic Entanglement Entropy,'' Lect.Notes Phys. {\bf 931} (2017)
  \href{http://arxiv.org/abs/1609.01287}{[arXiv:1609.01287[hep-th]]}.

\bibitem{mm2} 
 Parul Jain, Siddhi Swarupa Jena, Subhash Mahapatra,
  ``Holographic confining-deconfining gauge theories and entanglement measures with a magnetic field,'' Phys. Rev. D {\bf 107} (2023) 8, 086016,
  \href{http://arxiv.org/abs/2209.15355}{[arXiv:2209.15355[hep-th]]}.

\bibitem{mm3} 
 Roldao da Rocha,
  ``Holographic entanglement entropy, deformed black branes, and deconfinement in AdS/QCD,'' Phys. Rev. D {\bf 105} (2022) 2, 026014,
  \href{http://arxiv.org/abs/2111.01244}{[arXiv:2111.01244[hep-th]]}.

\bibitem{mm4} 
David Dudal, Subhash Mahapatra,
  ``Interplay between the holographic QCD phase diagram and entanglement entropy,'' JHEP {\bf 07} (2018) 120, 
  \href{http://arxiv.org/abs/1805.02938}{[arXiv:1805.02938[hep-th]]}.

\bibitem{mm5} 
Zhibin Li, Kun Xu, Mei Huang,
  ``The entanglement properties of holographic QCD model with a critical end point,'' Chin.Phys.C {\bf 45} (2021) 1, 013116, 
  \href{http://arxiv.org/abs/2002.08650}{[arXiv:2002.08650[hep-th]]}.

\bibitem{mm6} 
Irina Ya. Aref'eva, Alexander Patrushev, Pavel Slepov,
  ``Holographic entanglement entropy in anisotropic background with confinement-deconfinement phase transition,'' JHEP {\bf 07} (2020) 043, 
  \href{http://arxiv.org/abs/2003.05847}{[arXiv:2003.05847[hep-th]]}.


\bibitem{mm7} 
Parul Jain, Subhash Mahapatra,
  ``Mixed state entanglement measures as probe for confinement,'' Phys.Rev.D {\bf 102} (2020) 126022, 
  \href{http://arxiv.org/abs/2010.07702}{[arXiv:2010.07702[hep-th]]}.

\bibitem{mm8} 
M. Ali-Akbari, M. Lezgi,
  ``Holographic QCD, entanglement entropy, and critical temperature,'' Phys.Rev.D {\bf 96} (2017) 8, 086014, 
  \href{http://arxiv.org/abs/1706.04335}{[arXiv:1706.04335[hep-th]]}.
  
  \bibitem{mm9} 
M. Rahimi, M. Ali-Akbari, M. Lezgi,
  ``Entanglement entropy in a non-conformal background
,'' Phys.Lett.B {\bf 771} (2017) 583-587, 
  \href{http://arxiv.org/abs/1610.01835}{[arXiv:1610.01835[hep-th]]}.

\bibitem{cc1}
John Watrous, ``Quantum Computational Complexity,'' 
\href{http://arxiv.org/abs/0804.3401}{[arXiv:0804.3401[quant-ph]]}.

\bibitem{cc2}
 Micheal A. Nielsen and Isaac L. Chuang, ``Quantum Computation and Quantum Infromation,'' Cambridge university press, (2010), 702 p.


\bibitem{cc3}
R.~Jefferson and R.~C.~Myers ``Circuit complexity in quantum field theory,'' JHEP  \textbf{10} (2017) 107, 
\href{http://arxiv.org/abs/11707.08570}{[arXiv:1707.08570[quant-ph]]}.


\bibitem{susskind1}
D. Stanford and L. Susskind, ``Complexity and Shock Wave Geometries,'' Phys. Rev. D \textbf{90}, no.12, 126007 (2014) 
 \href{http://arxiv.org/abs/1406.2678}{[arXiv:1406.2678[hep-th]]}.
\bibitem{susskind2}
A. R. Brown, D. A. Roberts, L. Susskind, B. Swingle and Y. Zhao, ``Complexity, action, and
black holes,'' Phys. Rev. D \textbf{93}, no.8, 086006 (2016) \href{http://arxiv.org/abs/1512.04993}{[arXiv:1512.04993[hep-th]]}. 


\bibitem{comments} 
    Dean~Carmi, Robert~C. Myers and Pratik~Rath
  ``Comments on Holographic Complexity,'' JHEP {\bf 1703}, 118 (2017)
  \href{http://arxiv.org/abs/1612.00433}{[arXiv:1612.00433[hep-th]]}. 


  \bibitem{alishahiha}
  Mohsen~Alishahiha,
  ``Holographic complexity,'' Phys.~Rev.~D.~ {\bf 92}, 126009 (2015)
  \href{http://arxiv.org/abs/1509.06614}{[arXiv:1509.06614[hep-th]]}. 


\bibitem{v1} 
    Omer~Ben-Ami, Dean~Carmi,
  ``On Volumes of Subregions in Holography and Complexity,'' JHEP {\bf 1611}, 129 (2016)
  \href{http://arxiv.org/abs/1609.02514}{[arXiv:1609.02514[hep-th]]}.

\bibitem{v2} 
  S.~J.~Zhang,
  ``Complexity and phase transitions in a holographic QCD model,''
  Nucl.\ Phys.\ B {\bf 929}, 243 (2018)
  \href{http://arxiv.org/abs/1712.07583}{[arXiv:1712.07583[hep-th]]}.
 
\bibitem{v3} 
  S.~J.~Zhang,
  ``Subregion complexity in holographic thermalization with dS boundary,'' Eur.Phys.J.C {\bf 79} (2019) 8,715
  \href{http://arxiv.org/abs/1905.10605}{[arXiv:1905.10605[hep-th]]}.

\bibitem{v4} 
  Pratim~Roy, Tapobrata~Sarkar, 
  ``On subregion holographic complexity and renormalization group flows,'' Phys.Rev. D {\bf 97}, 086018 (2018)
  \href{http://arxiv.org/abs/1708.05313}{[arXiv:1708.05313[hep-th]]}.
 
\bibitem{v5} 
  R Fareghbal and P Karimi, 
  ``Complexity growth in flat spacetimes,''
  Phys. Rev. D {\bf 98}, no. 4, 046003 (2018)
  \href{http://arxiv.org/abs/1806.07273}{[arXiv:1806.07273[hep-th]]}.

\bibitem{v6} 
 M.Alishahiha, A.Faraji Astaneh, M.R.Mohammadi Mozaffar and A.Mollabashi, 
   ``Complexity Growth with Lifshitz Scaling and Hyperscaling Violation,''
  JHEP {\bf 1807}, 042 (2018)
  \href{http://arxiv.org/abs/1802.06740 }{[arXiv:1802.06740 [hep-th]]}.
  
 
\bibitem{v7} 
M.~Alishahiha, K.~Babaei Velni and M.~R.~Mohammadi Mozaffar,
  ``Subregion Action and Complexity,'' Phys.Rev.D {\bf 99} (2019) 12, 126016,
\href{http://arxiv.org/abs/1809.06031 }{[arXiv:1809.06031 [hep-th]]}.
 

\bibitem{v8} 
 Mahsa~Lezgi and Mohammad~Ali-Akbari
 ``A note on holographic subregion complexity and QCD phase transition, "
 Phys. Rev. D {\bf 101}, 026022 (2020)
  \href{http://arxiv.org/abs/1908.01303}{[arXiv:1908.01303[hep-th]]}. 

\bibitem{v9} 
 M.~asadi,
 ``On volume subregion complexity in non-conformal theories,"
 Eur.Phys.J.C 80 (2020) 7, 681
  \href{http://arxiv.org/abs/2004.11306}{[arXiv:2004.11306[hep-th]]}.


\bibitem{v10} 
 Mahsa Lezgi, Mohammad Ali-Akbari and Mohammad Asadi,
 ``Non-Conformality, Subregion Complexity and Meson Binding"
  Phys.Rev.D {\bf 104} (2021) 2, 026001,
  \href{http://arxiv.org/abs/2011.11625}{[arXiv:2011.11625[hep-th]]}.

\bibitem{v11}
  Mahsa lezgi and Mohammad Ali-Akbari,
  ``Complexity and uncomplexity during energy injection,'' 
  Phys.Rev.D {\bf 103} (2021) 12, 126024
  \href{http://arxiv.org/abs/2103.05023}{[arXiv:2103.05023[hep-th]]}.

\bibitem{v12}
 Mohammad Ali-Akbari and Mahsa Lezgi,
  ``Note on stability and holographic subregion complexity,'' 
  Eur.Phys.J.C {\bf 82} (2022) 2, 114,
  \href{http://arxiv.org/abs/2110.05793}{[arXiv:2110.05793[hep-th]]}.
  
  \bibitem{v13}
   Mohammad Ali-Akbari and Mahsa Lezgi,
  ``Resource and stability near a critical point from the quantum information perspective,'' 
  Phys.Lett.B {\bf 842} (2023) 137954,
  \href{http://arxiv.org/abs/2209.04623}{[arXiv:2209.04623[hep-th]]}.

\bibitem{v14}
  Mohammad Ali-Akbari and Mahsa Lezgi,
  ``Subregion volume complexity under thermal and electromagnetic quenches,'' Phys.Rev.D, {\bf 108},(2023)8, 086023,
   \href{http://arxiv.org/abs/2308.04900}{[arXiv:2308.04900[hep-th]]}.

\bibitem{v15} 
Andrew R. Frey, Michael P. Grehan, Manu Srivastava,
 ``Complexity of scalar collapse in anti-de Sitter spacetime"
JHEP {\bf 12} (2021) 135,
  \href{http://arxiv.org/abs/2110.09630}{[arXiv:2110.09630[hep-th]]}.

\bibitem{hotwind}
 Hong Liu,  Krishna Rajagopal, Urs Achim Wiedemann, 
  ``An AdS/CFT Calculation of Screening in a Hot Wind,'' Phys.Rev.Lett. {\bf 98} (2007) 182301,
   \href{http://arxiv.org/abs/0607062}{[arXiv:0607062[hep-ph]]}. 

\bibitem{AliAkabri}
 M. Ali-Akbari, D. Giataganas, Z. Rezaei, 
  ``Imaginary potential of heavy quarkonia moving in strongly coupled plasma,'' Phys.Rev.D {\bf 90} (2014)8, 086001,
   \href{http://arxiv.org/abs/1406.1994}{[arXiv:1406.1994[hep-ph]]}. 

\bibitem{boost4}
 David D. Blanco, Horacio Casini,  Ling-Yan Hung,  Robert C. Myers,
  ``Relative Entropy and Holography,'' JHEP {\bf 08} (2013) 060,
   \href{http://arxiv.org/abs/1305.3182}{[arXiv:1305.3182[hep-th]]}.

\bibitem{boost1}
 Rohit Mishra,  Harvendra Singh, 
  ``Entanglement asymmetry for boosted black branes and the bound,'' Int.J.Mod.Phys.A {\bf 32} (2017) 16,
   \href{http://arxiv.org/abs/1603.06058}{[arXiv:1603.06058[hep-th]]}.
   
\bibitem{boost2}
 Sourav Karar,  Rohit Mishra, Sunandan Gangopadhyay 
  ``Holographic complexity of boosted black brane and Fisher information,'' Phys.Rev.D  {\bf 100} (2019) 2,026006,
   \href{http://arxiv.org/abs/1904.13090}{[arXiv:1904.13090[hep-th]]}.
   
\bibitem{boost3}
 Anirban Chowdhury Roy, Ashis Saha, Sunandan Gangopadhyay
  ``Mixed state information theoretic measures in boosted black brane,'' Annals Phys. {\bf 452} (2023) 169270,
   \href{http://arxiv.org/abs/2204.08012}{[arXiv:2204.08012[hep-th]]}.

\bibitem{boost5}
 Atanu Bhatta, Shankhadeep Chakrabortty, Suat Dengiz,  Ercan Kilicarslan,
  ``High temperature behavior of non-local observables in boosted strongly coupled plasma: A holographic study,'' Eur.Phys.J.C {\bf 80} (2020) 7, 663,
   \href{http://arxiv.org/abs/1909.03088}{[arXiv:1909.03088[hep-th]]}.


\bibitem{metric}
 Jyotirmoy Bhattacharya, Parthajit Biswas,  A. Chandranathan, Sayan Kumar Das,
  ``Holographic entanglement entropy for relativistic hydrodynamic flows,'' JHEP {\bf 05} (2023) 092,
   \href{http://arxiv.org/abs/2211.14271}{[arXiv:2211.14271[hep-th]]}.
   
\bibitem{mixed} 
Cesar A. Agon, Santa Barbara, Matthew Headrick, Brian Swingle,
 ``Subsystem Complexity and Holography"
JHEP {\bf 02} (2019) 145,
  \href{http://arxiv.org/abs/1804.01561}{[arXiv:1804.01561[hep-th]]}.    
   

\bibitem{A1}
  Matthew~N.O.~Sadiku,
  ``Numerical Techniques in Electromagnetics,'' Routledge publishing, (2000).
  
\bibitem{vad}
 Roberto Auzzi, Giuseppe Nardelli, Fidel l. Schaposnik Massolo, Gianni Tallarita, Nicolo Zenoni,
  ``On volume subregion complexity in Vaidya spacetime,'' JHEP {\bf 11} (2019) 098,
   \href{http://arxiv.org/abs/1908.10832}{[arXiv:1908.10832[hep-th]]}.

\end{thebibliography}
\end{document}